\documentclass{article}

\usepackage{PRIMEarxiv}

\usepackage[utf8]{inputenc} 
\usepackage[T1]{fontenc}    
\usepackage{hyperref}       
\usepackage{url}            
\usepackage{booktabs}       
\usepackage{amsfonts}       
\usepackage{nicefrac}       
\usepackage{microtype}      
\usepackage{lipsum}
\usepackage{fancyhdr}       
\usepackage{graphicx}       
\graphicspath{{media/}}     

\usepackage[numbers]{natbib}
\usepackage{amsmath}
\usepackage{listings}

\title{Numerical studies of triangulated vesicles with anisotropic membrane inclusions
}

\author{
  Yoav Ravid, Nir Gov \\
  Department of Chemical and Biological Physics \\
  Weizmann Institute of Science \\
  Rehovot 7610001, Israel\\
\\
   \And
  Mitja Drab, Aleš Iglič \\
  Laboratory of Physics \\
  Faculty of Electrical Engineering \\
 University of Ljubljana, 1000 Ljubljana, Slovenia\\
  \texttt{Corresponding author: mitja.drab@fe.uni-lj.si} \\
   \AND
   Veronika Kralj-Iglič \\
  Laboratory of Clinical Biophysics \\
  Faculty of Health Sciences \\
 University of Ljubljana, 1000 Ljubljana, Slovenia\\
 \And
   Samo Penič \\
  Laboratory of Bioelectromagnetics\\
  Faculty of Electrical Engineering \\
 University of Ljubljana, 1000 Ljubljana, Slovenia\\
}

\begin{document}
\maketitle

\begin{abstract}
In this study, we implement the deviatoric curvature model to examine dynamically triangulated surfaces with anisotropic membrane inclusions. The Monte-Carlo numerical scheme is devised to not only minimize the total bending energy of the membrane but also the in-plane nematic order of the inclusions by considering the mismatch between the curvature of the membrane and the intrinsic curvature of the inclusion. Neighboring inclusions can either attract with nearest-neighbor interaction or with a nematic interaction derived from liquid crystal theory. Orientational order determines whether vesicles fully covered with inclusions result in bulbs connected by necks or long tubes. Remarkably, when inclusions on vesicles with no vacancies interact non-nematically, a spontaneous local order can lead to a bulb transition which may have implications in cell or organelle division. Furthermore we find that average nematic order is inversely proportional to the number of thin necks formed in the vesicles. Our method shows good convergence and is suitable for further upgrades, for example to vesicles constrained by volume.
\end{abstract}


\section{Introduction}

Cellular shape holds immense importance in various  functions like division, movement, and signaling. In biophysics, there's a significant interest in comprehending how cells take shape. Research involving theories and computer simulations indicates that the physical traits of the cell membrane, like its uneven components and dynamic forces, significantly influence structure. The precise interplay between these elements in shaping cells remains uncertain.

Membrane shape depends on the intrinsic shape of the membrane's molecular constituents \cite{israelachvili2022surface,kralj2002deviatoric,kralj1996shapes,kralj2012stability,fournier1996nontopological} and their interactions with other components, such as membrane skeleton and cytoskeleton \cite{markin81, leibler86,kig99, iglJBiomech2007,Fosnaric06,fosnaric2019theoretical,gov2018guided, hagerstrand2006curvature,iglivc1997possible,mahapatra2021mechanics,alimohamadi2021mechanical}. It has been shown that a non-homogeneous lateral distribution and phase separation of membrane inclusions may be a driving force of cell shape transformations and necessary for the stabilization of highly curved
membrane structures \cite{ iglicRBC2007,veksler2007phase,bozic2006,drab2019inception,schamberger2023curvature,mcmahon2005membrane,zimmerberg2006proteins,kralj2002deviatoric,kralj1996shapes}.

In recent years, there have been significant advances in experimental techniques for studying cell mechanics \cite{drab2021monte,disanza2013cdc42,pandur2023surfactin}. However, theoretical and computational studies remain important for gaining insight into the underlying physical mechanisms. For example, the study of vesicle shapes driven by coupling curved inclusions and active cytoskeletal forces has shown that the inhomogeneous nature of membrane constituents, such as curved inclusions, can have a significant impact on vesicle shapes \cite{fosnaric2019theoretical,drab2019inception,drab2023modeling,takatori2020active,mancinelli2021dendrite,cagnetta2022universal,shrestha2021mechanics,maji2023shape,stachowiak2021membrane}. 

In this work, we apply the deviatoric curvature model to investigate dynamically triangulated surfaces containing anisotropic membrane inclusions that are banana shaped. A Monte-Carlo numerical approach is employed not only to reduce the overall bending energy of the membrane but also to minimize the in-plane nematic order of the inclusions. This is achieved by considering the discrepancy between the curvature of the membrane and the inclusion it contains. We find that the orientational order has an effect on the overall limiting cell shape. Additionally, we present a possible mechanism of cell division when inclusions interact non-nematically through spontaneous ordering. Furthermore, we find that the total bonding energy of the fully covered vesicles is inversely proportional to the number of necks and buds in the vesicles. 

\section{Theoretical background}
\label{sec:headings}

Theoretical background is based on  the derivation of membrane energy and anisotropic inclusion interactions, as outlined by Kralj-Iglič et al. \cite{iglivc2005role,kralj2002deviatoric,kralj2012stability}. The model presents the membrane as a 2D anisotropic continuum surface. The elastic energy of a small membrane section becomes null when its principal curvatures $C_1$ and $C_2$ align with their intrinsic counterparts of the anisotropic inclusions $C_{1m}$ and $C_{2m}$, and when the orientations of the local membrane curvature tensor $C$ coincide with the intrinsic membrane curvature tensor $C_m$. A shape with these properties throughout exhibits zero elastic energy.

\begin{figure}
    \centering
    \includegraphics[width=.6\textwidth]{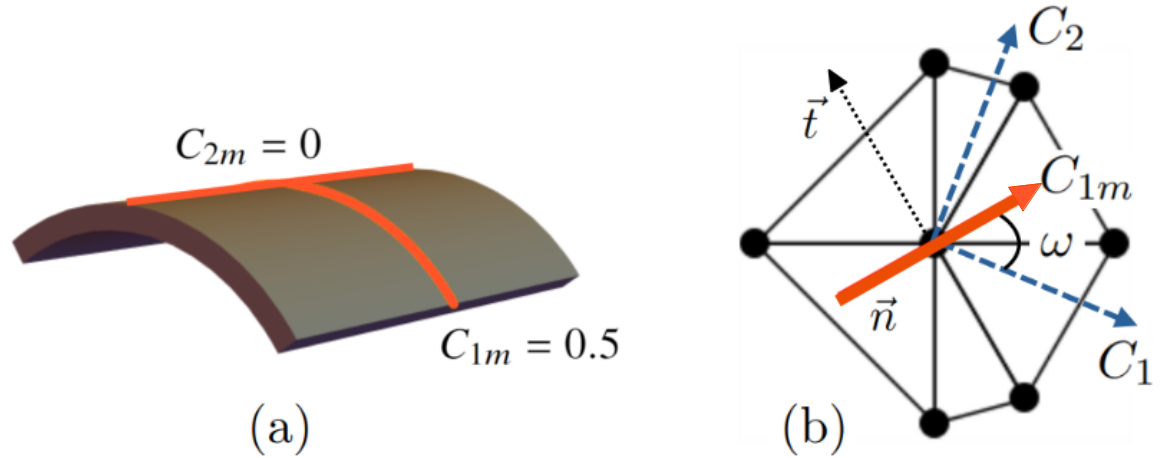}
    \caption{Example of a banana inclusion. For this particular inclusion, $H_m=D_m=0.25$ (a). The mismatch between a membrane inclusion and local curvature of the triangulated membrane (b). Here, $\vec{n}$ is the orientation of the inclusion and $\vec{t}$ is its perpendicular direction in the local tangent plane.}
    \label{fig:fig1}
\end{figure}

In essence, the elastic energy per unit area of a thin plate element with area $dA$ is defined by the discrepancy between its actual local membrane curvature and its intrinsic curvature. Representing the actual and intrinsic curvatures with tensors $C$ and $C_m$, respectively, where in their principal systems, the curvature tensor matrices feature only diagonal elements \cite{kralj2002deviatoric,iglivc2005role}:
$$
C = \begin{pmatrix} C_1 & 0 \\ 0 & C_2 \end{pmatrix}, \quad C_m = \begin{pmatrix} C_{1m} & 0 \\ 0 & C_{2m} \end{pmatrix}.
$$
In this examination, the principal systems of tensors generally undergo an angle $\omega$ rotation in the tangent plane. The disparity between the actual and intrinsic local membrane curvature is expressed by the tensor \cite{kralj2002deviatoric,kralj2012stability} $M = R C_m R^{-1} - C$, involving the rotation matrix
$$
R = \begin{pmatrix} \cos \omega & -\sin \omega \\  \sin \omega & \cos \omega \end{pmatrix}.
$$
A small shell patch adjusts to conform to the actual membrane, reflecting the energy required for deformation (see Figure \ref{fig:fig1}(b)). The approximate elastic energy $E_1$ unfolds through an expansion in powers of all independent invariants of the tensor $M$. Employing the trace and determinant of the tensor as invariants yields \cite{iglivc2005role,kralj2002deviatoric,kralj2012stability,kralj1996shapes}:
\begin{equation}\label{eq:eq1}
    E_1 = \int \frac{K_1}{2} (\mathrm{Tr} M)^2 + K_2 \mathrm{Det} M \,dA,
\end{equation}
integrated across the entire vesicle. With constants $K_1$ and $K_2$, the energy $E_1$ is \cite{iglivc2005role,kralj2002deviatoric,kralj2012stability,kralj1999free}:
\begin{equation}\label{eq:eq2}
E_1 = \int (2K_1 + K_2)(H - H_m)^2 - K_2 \left( D^2 - 2DD_m \cos 2\omega + D_m^2 \right)\,dA,
\end{equation}
where $D = (C_1 - C_2)/2$ stands as the curvature deviator, an invariant of the curvature tensor ($D^2 = (\mathrm{Tr}(C)/2)^2 - \mathrm{Det}(C) = H^2 - C_1C_2$); $H_m = (C_{1m} + C_{2m})/2$ denotes the intrinsic mean curvature; and $D_m = (C_{1m} - C_{2m})/2$ represents the intrinsic curvature deviator.

In the case of an isotropic membrane where $D_m=0$, it becomes evident that this equates to the Helfrich bending energy density \cite{helfrich1973} described by $E_b = k^2_c (2H - C_0)^2 + k_G K$, with $H = (C_1 + C_2)/2$ as the mean curvature, $C_0$ as the spontaneous curvature, $K = C_1C_2$ as the Gaussian curvature, and $k_c$ and $k_G$ as the bending and splay modulus, respectively \cite{iglivc2005role,kralj1996shapes,kralj2012stability}.

\subsection{Binding energy}

Inclusions may bond on the membrane by a simple energy gain for two neighboring vertices.
\begin{equation}\label{eq:eq3}
    E_2 = \begin{cases} - w & i,j\text{ are binding} \\ 0 & \text{else} \end{cases}
\end{equation}
where the type and the binding strength $w$ are vertex properties, reflecting aggregation. Throughout this paper, we refer to this type of bonding as normal bonding. This kind of bonding is observed in surfactants and proteins, since it arises from spontaneous self-assembly on the membrane \cite{deleu2003interaction,fosnaric2019theoretical}.   

\subsection{Nematic energy}

If we want to impose a directional energy penalty to neighboring inclusions, we use a nematic-nematic interaction employed from liquid crystal theory. The energy between neighboring inclusions will be lower if their principal curvatures are aligned. We model this based on Frank's free energy density for nematic liquid crystals \cite{frank1958liquid}:
\begin{equation}\label{eq:eq4}
\begin{aligned}
E_2 &=  \int \left(\frac{k_G}{2}\left(\nabla\cdot\vec{n}\right)^2 + \frac{k_c}{2}\left(\nabla\cdot\vec{t}\right)^2\right) r_v \,dA
\end{aligned}
\end{equation}
Here, $\nabla$ represents the covariant derivative on the curved surface, $\vec{n}$ denotes the orientation of the inclusion, and $\vec{t}$ signifies its perpendicular direction in the local tangent plane. The constants $k_G$ and $k_c$ correspond to the splay and bending elastic constants governing in-plane nematic interactions. The variable $r_v$ takes a value of 1 if a nematic is present on a vertex; otherwise, it is 0.

A discrete form of this energy is employed to facilitate implementation in Monte-Carlo (MC) simulations. If we assume a one-constant approximation ($k_G=k_c$), equation \ref{eq:eq4} can be rewritten in a way that makes the implementation suitable for MC simulations (Lebwohl-Lasher model) \cite{lasher1972monte}:
\begin{equation}\label{eq:eq5}
    E_2 \approx - \int \sum_{k=1}^{N} \epsilon_{LL}^{k,k} \sum_{i>j}\left(\dfrac{3}{2}(\vec{n_i} \cdot \vec{n_j})^2 -\dfrac{1}{2} \right)r_v \,dA. 
\end{equation}
Here, $\epsilon_{LL}$ is the strength of the nematic interaction. The sum $\sum_{i>j}$ is over all the nearest neighbor (i, j) vertices on the triangulated grid, promoting alignment among the neighbouring orientation vectors. An even simpler form of this approximation for the in-plane orientational field is the XY model on a random surface \cite{ramakrishnan2010monte}:
\begin{equation}\label{eq:eq6}
E_2 \approx -w \int \sum_{i>j}(\vec{n}_i\cdot \vec{n}_j)^2 \, dA.
\end{equation}\label{eq:nematic}

Here, $w$ represents the strength of the direct interaction constant, and the summation runs over all inclusion-inclusion pairs. The sum of $E_1$ and $E_2$ is minimized numerically, while $E_2$ is either normal (equation \ref{eq:eq3}) or nematic bonding  (equation \ref{eq:eq6}). There is no volume restricting the vesicle shapes. All the inclusions in this work are banana-shaped (see Figure 1.1(a)). Normal interaction reflects the tendency of inclusions to self-aggregate. Nematic interaction goes further and implies the tendency to both self-aggregate and also align inclusions' principal curvatures.

The details of the Monte-Carlo procedure are given in Appendix \ref{app:procedure} and the details of the mesh generation and finding the principal curvatures are given in in Appendix \ref{app:code details}.

\section{Results and discussion}\label{sec3}
\subsection{Small meshes}

\begin{figure}[ht]
\centering
  \includegraphics[width=0.8\textwidth]{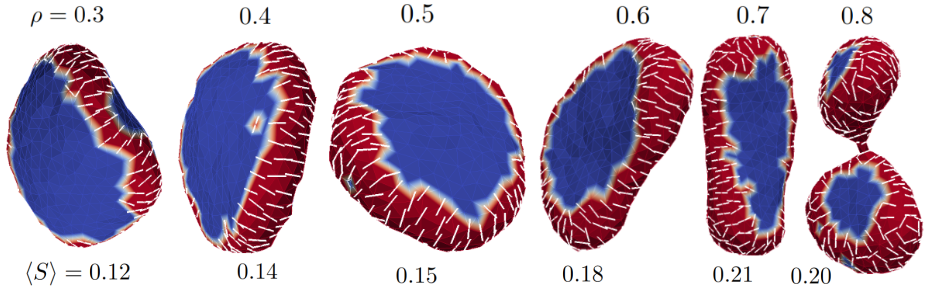}
  \caption{A sequence of equilibrium shapes for a small mesh with normal interaction between banana inclusions. The constant of inclusion interaction is $w=3$, while the curvatures are $H_m=D_m=0.25$. The inclusions are marked in red, the empty membrane in blue. The principal direction of the curvature $C_{1m}$ is shown with the white lines. The average order parameter of each shape $\langle S \rangle$ is given in the bottom and calculated by equation \ref{eq:eq8}}\label{fig:fig2}
\end{figure}

We start by examining smaller vesicles ($N=502$ vertices) with a fraction of their surface $\rho$ covered with banana shaped inclusions as shown in Figure \ref{fig:fig1}(a). One of their main curvatures is positive and another is zero, which translates to $H_m=D_m=0.25$. With increasing $\rho$ and normal interaction between inclusions, we get the sequence shown in Figure \ref{fig:fig2}. Due to the inclusions' strong affinity to aggregate ($w=3$), they form a rim around the perimeter of the vesicle even at low values of $\rho$. With increasing $\rho$, the vesicles become flattened and oblate and later elongated (but not prolate). Only above $\rho=0.7$ does the high inclusion content clip the vesicle into two bulbs connected by a thin neck. 

We want to quantify the degree of order between neighboring anisotropic inclusions. For this reason we can define the nematic order of the inclusion at vertex $i$ by
\begin{equation}\label{eq:eq7}
    S_i = \frac{1}{2}(3\cos^2{\theta}-1),
\end{equation}
where the angle $\theta$ measures the angle between directors (or principal curvature $C_{1m}$) of neighboring inclusions on the membrane. The more aligned the neighboring inclusions are, the more this value approaches unity. In our numerical approach it is calculated as follows: at vertex $i$, we average the dot product of the director $\vec{d}_i$ with all immediate neighboring vertices (see Appendix \ref{app:order}). The average nematic order across the entire membrane is then
\begin{equation}\label{eq:eq8}
    \langle S \rangle = \frac{1}{N}\sum_{i=1}^N S_i.
\end{equation}
Returning to Figure \ref{fig:fig2}, we notice that predictably the nematic order increases with increasing $\rho$, but peaks just before the bulb transition. Their binding promotes a rim on the membrane that accomodates their spontaneous curvature best when they all align in a similar direction (see Figure \ref{fig:fig2} for $\rho=$ 0.3--0.7). When the inclusion density is large enough, the ordering ceases, resulting in two prolate bulbs connected by a neck where nematic order decreases. We call this the bulb transition. The average order parameter at the bottom of Figure 1.2. shows that the bulb transition decreases the average order.

\begin{figure}[ht]
    \centering
    \includegraphics[width=0.65\textwidth]{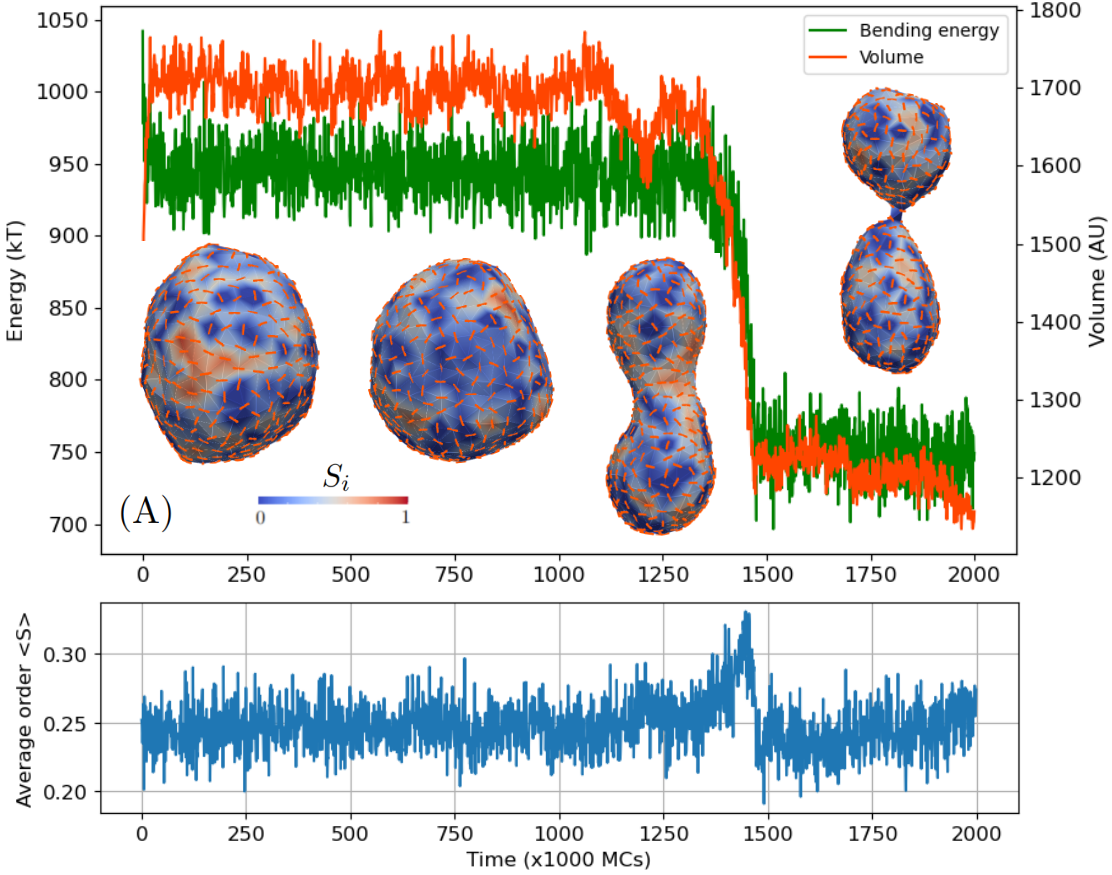}
    \includegraphics[width=0.65\textwidth]{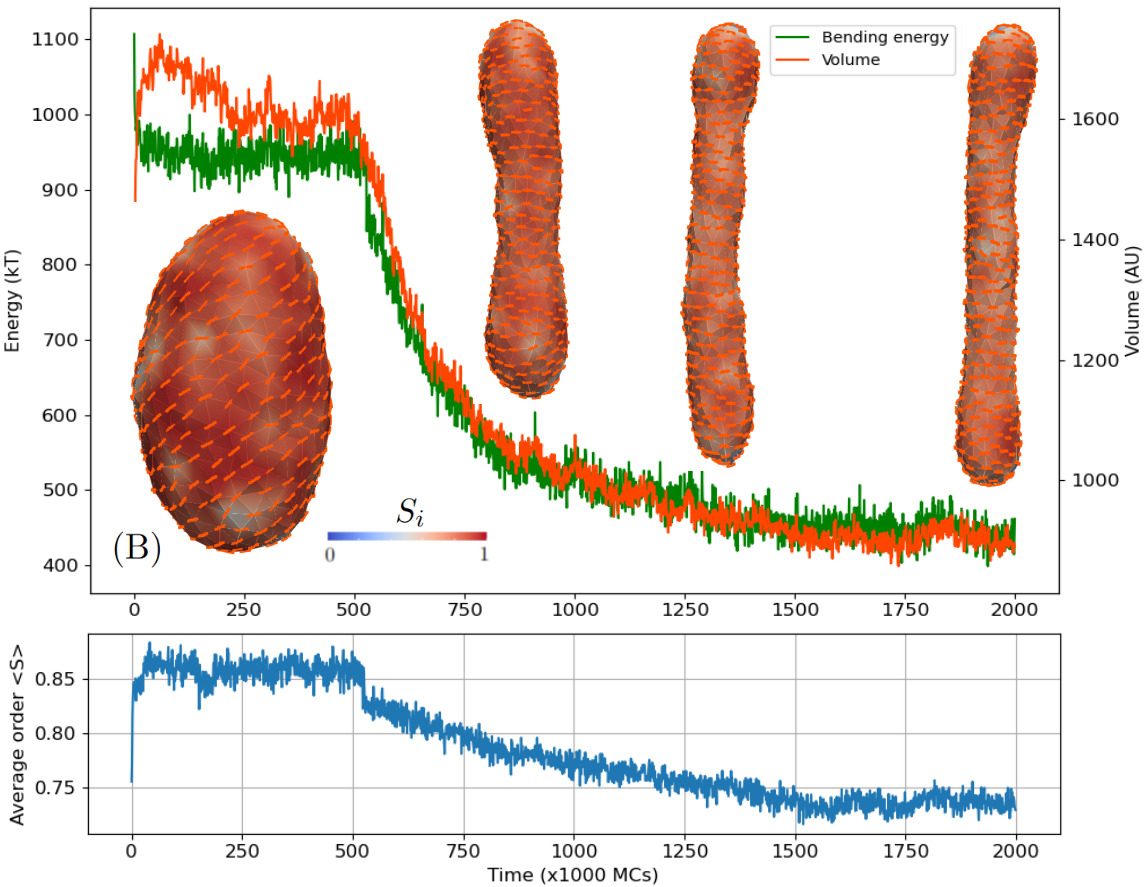}
    \caption{Simulations of vesicles fully covered in banana inclusions ($H_m=D_m=0.25$) for a small mesh (N=502) and normal interaction ($w=3$)(A) and nematic interaction (B). With consequent MC steps, the shapes relax to an energy minimum. For normal interactions, the resulting shape is a two bulbs connected by a thin neck. For the nematic case, the limiting shape is an elongated and thin cylinder. The average order parameter $\langle S \rangle$ for both cases is shown below the main graph.}
    \label{fig:fig3}
\end{figure}

Next we investigate vesicles with all places on the membrane occupied by inclusions. These two limiting cases of $\rho=1$ are shown in Figure \ref{fig:fig3}. Here, we observe the evolution of the vesicle shapes from its spherical start. We may think of consequential MC steps as simulation time shown on the $x$-axis. 

First, we look at the normal interaction between inclusions shown in  Figure \ref{fig:fig3}(A). The inclusions predictably show no sign of nematic order on the membrane, prompting neighbours to point in a random direction and hence promote the spherical shape that persists for a long time. This is because the average curvature of many random directions results in a net zero curvature and a spherical shape. However, the fluctuations with time lead again to a bulb transition.
The bulb transition coincides with a sharp decline in the bending energy (accompanied by a decline in volume) shown in Figure \ref{fig:fig3}(A). The average nematic order slightly increases before it returns to its pre-transition values. 

\begin{figure}[ht]
    \centering
    \includegraphics[width=0.5\textwidth]{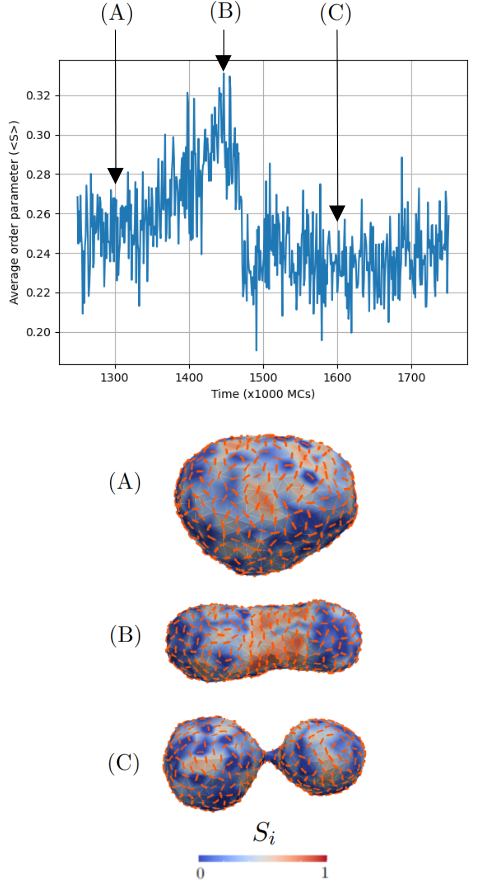}
    \caption{A close-up of the average order parameter during the bulb transition (see Figure \ref{fig:fig3}(A)). A spontaneous ordering of the inclusions results in the transition from a sphere-like shape (A) to bulbs connected by a thin neck (B). After the transition, $\langle S \rangle$ decreases together with the bending energy.}
    \label{fig:fig4}
\end{figure}

Upon closer inspection of an increase in $\langle S \rangle$ preceding the bulb transition for Figure \ref{fig:fig3}(A), we notice an interesting sequence of events shown in Figure \ref{fig:fig4}. Inclusions spontaneously order along the circumference of the prolate vesicle, resulting in an increase of $\langle S \rangle$ (see Figure \ref{fig:fig4}(B)). This is followed by a sudden narrowing of the midsection and results in a bulb transition (Figure \ref{fig:fig4}(C)). Here, $\langle S \rangle$ and bending energy decrease relative to their pre-transition values. The spontaneous ordering of the inclusions can therefore have lasting impact on the morphology of the vesicle and may facilitate in the mechanism of cell division.   

If we run the same simulation with nematic interactions between inclusions, the limiting shape is an elongated cylinder shown in Figure \ref{fig:fig3}(B). The bending energy continuously decreases in proportion with the average radius of the cylinder and leveling out at a limiting value. The overall bending energy of the thin cylinder is lower than the bulbs connected by a thin neck. 

\begin{figure}[ht]
    \centering
    \includegraphics[width=0.35\textwidth]{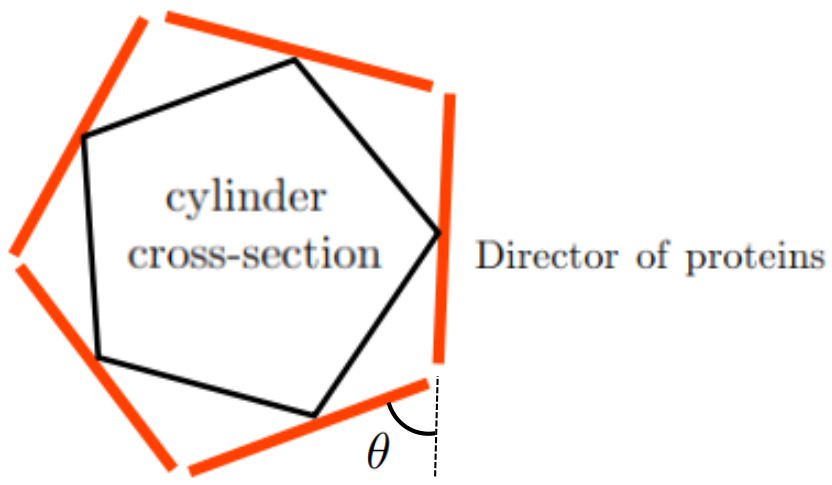}
    \caption{Low nematic order as a consequence of very thin necks of meshes. Although the curvature of the inclusions fits well around the circumference, the calculation of average nematic order given in equation \ref{eq:eq7} will be low because of the large angles ($\theta$) between neighboring inclusions. This artefact also accounts for near zero nematic order in the necks of buds.}
    \label{fig:fig5} 
\end{figure}

The nematic order interestingly slightly decreases (cca. 10\%) as the cylinder becomes elongated. This decrease may be a result of inclusions that are packed around a tighter radius. Let us envision this with a graphical example. In the limit of a very thin cylinder there may be only 5 vertices encircling the  smaller cylinder as shown in Figure \ref{fig:fig5}. Even though the inclusions' directions are aligned, neighboring angles are relatively large resulting in near zero nematic order (see Figure \ref{fig:fig5}).

\subsection{Larger meshes}

\begin{figure}[ht]
    \centering
    \includegraphics[width=0.5\textwidth]{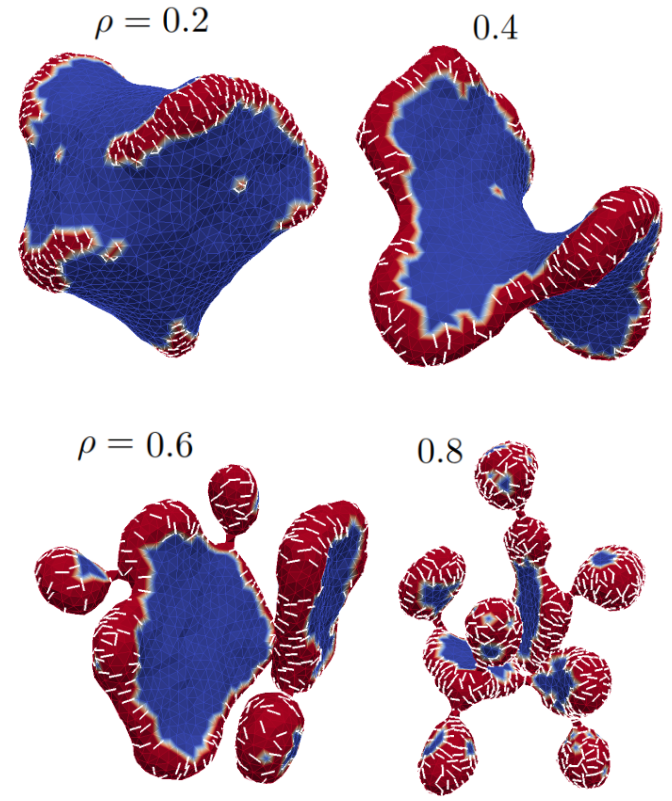}
    \caption{Simulation results for larger meshes ($N=2002$), $H_m=D_m=0.25$, normal interaction ($w=3$) after $2 \cdot 10^6$ MC steps in dependence of inclusion coverage fraction $\rho$. The color code is the same as in Figure \ref{fig:fig2}.}
    \label{fig:fig6}
\end{figure}

Now we turn our attention to larger vesicles ($N=2002$). We start again by gradually increasing the fraction of inclusions in the membrane with other parameters staying the same as before. The results are given in Figure \ref{fig:fig6}.

At $\rho=0.2$, the equilibrium shapes resemble quasi-spherical vesicles with isolated patches of inclusions forming ridges along the membrane surface. The ridges are formed due to the accommodation of the principal curvature of the inclusions $C_{1m}$. Increasing $\rho$ results in patches joining together to form a single ridge around the outer circumference of a flattened vesicle. As we can see in Figure \ref{fig:fig6} for $\rho=0.4$, the continuous ridge forces the vesicle shape away from the oblate pancake into a global saddle shape. Increasing $\rho$ further results in separation of a single saddle pancake to two separate ones. With $\rho=0.6$, each of the pancakes starts forming buds connected by a very thin neck. At $\rho=0.8$, nearly the whole surface of the mesh consists of buds covered with inclusions to resemble a pearling shape.  

\begin{figure}[ht]
    \centering
    \includegraphics[width=0.5\textwidth]{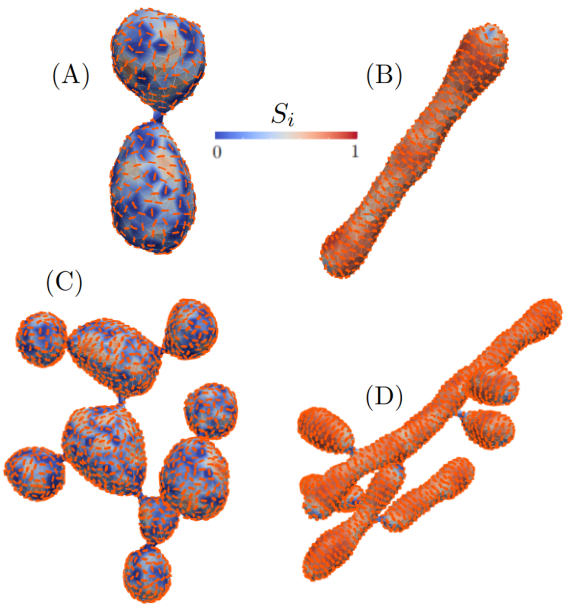}
    \caption{Nematic order as a heat map for small (A, B) and larger meshes (C, D). When the inclusion interaction is normal (A, C), there is little neighbor ordering shown by values close to 0 and a blue color. Conversely, red color represents strong nematic order as expected by the nematic neighbor interaction (B, D). Some spontaneous ordering may happen as seen by the orange patch in (A). The heatmap $S_i$ gives the local order parameter.}
    \label{fig:fig7}
\end{figure}

Figure \ref{fig:fig7} shows a heat map of the nematic order for a fully covered vesicle for a small (A, B) and larger mesh (C, D). Normal and nematic interaction between inclusions is considered for (A, C) and (B, D), respectively. We observe that nematic ordering results in elongated cylinders with nematic order near unity everywhere except at the poles of the cylinders, where a nematic defect is observed in blue (Figure \ref{fig:fig7}(B)). For larger meshes with nematic interactions the budding from the main trunk of the vesicle still occur, but the buds themselves are cylindrical. As expected, we notice that the nematic order is close to 0 in the necks where the radius is very thin. 

\begin{figure}[ht]
\centering
\begin{minipage}{0.5\textwidth}
  \centering
  \includegraphics[width=\linewidth]{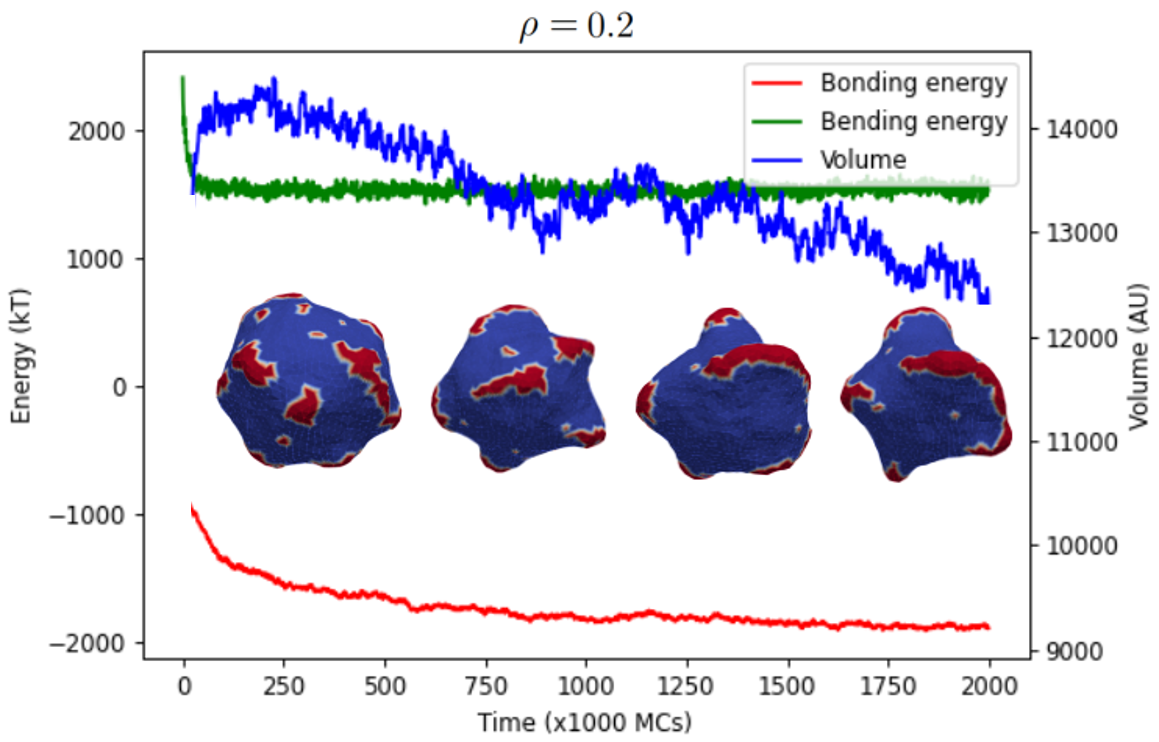}
  \caption{}
  \label{fig:fig8}
\end{minipage}%
\begin{minipage}{0.5\textwidth}
  \centering
  \includegraphics[width=\linewidth]{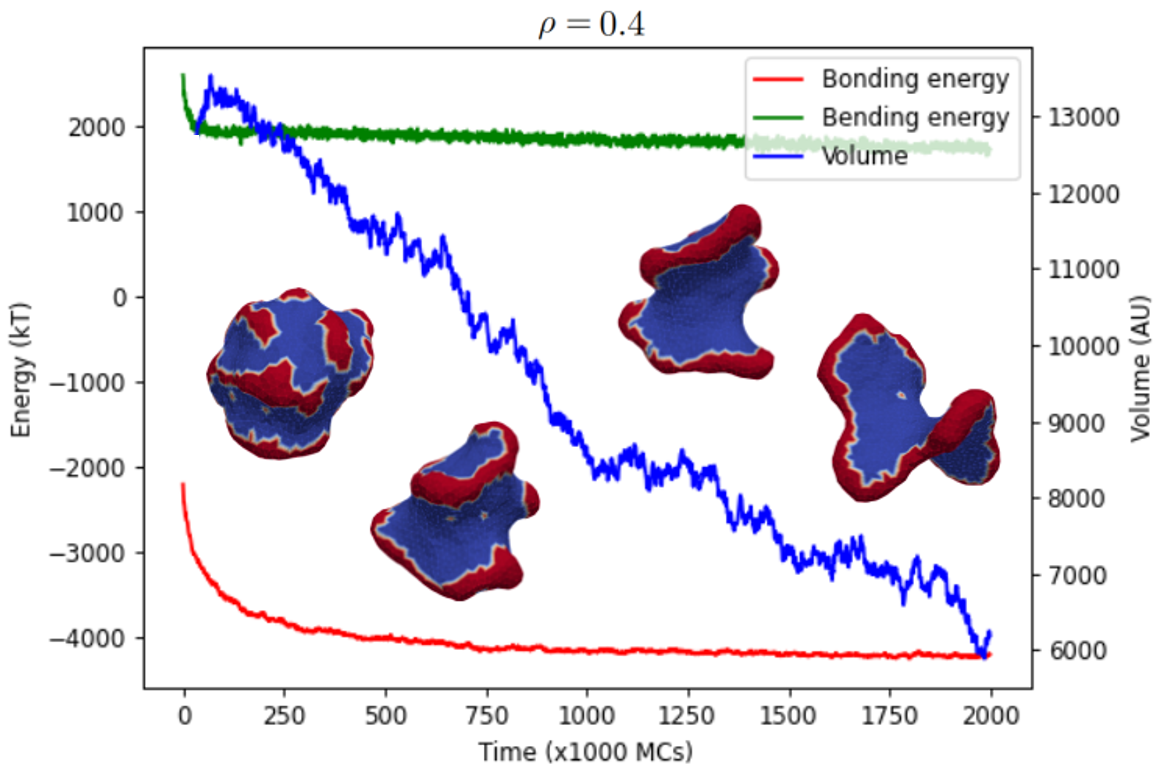}
  \caption{}
  \label{fig:fig9}
\end{minipage}

\begin{minipage}{0.5\textwidth}
  \centering
  \includegraphics[width=\linewidth]{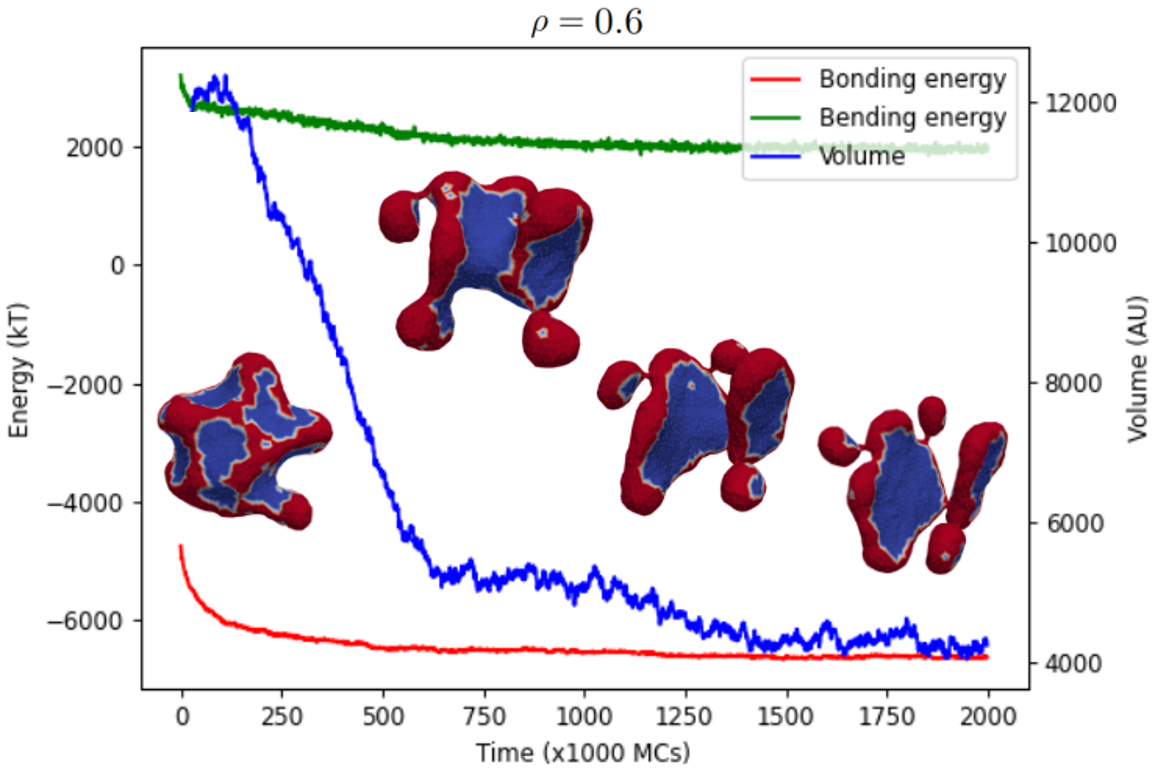}
  \caption{}
  \label{fig:fig10}
\end{minipage}%
\begin{minipage}{0.5\textwidth}
  \centering
  \includegraphics[width=\linewidth]{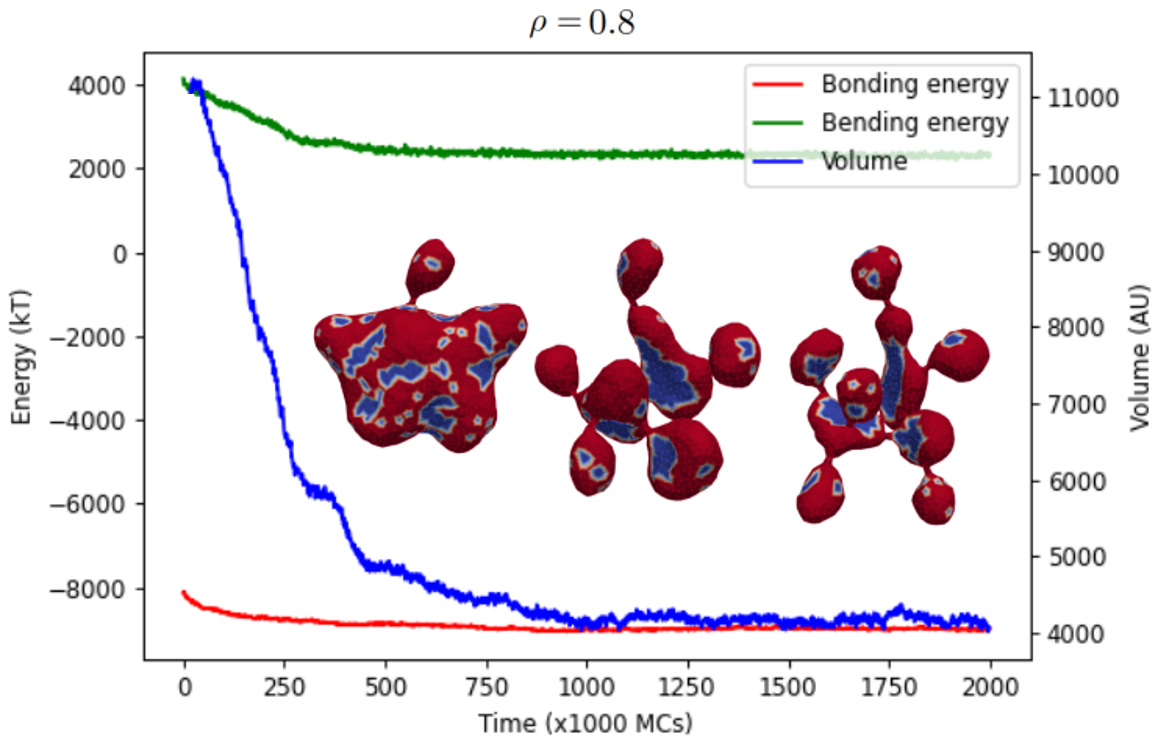}
  \caption{}
  \label{fig:fig11}
\end{minipage}
\end{figure}

For each final shape at constant $\rho$ for Figures \ref{fig:fig6}, Figures \ref{fig:fig8}--\ref{fig:fig11} show the bending energy, bonding energy and volume as a function of MC steps of the simulation. We observe that the vesicles approach their final volume faster in proportion to $\rho$. This likely happens as a consequence of budding which is more prevalent at higher inclusion content. The limiting pearling shape has the least volume per given vesicle area and the largest bending energy of all the shapes in the sequence of Figure \ref{fig:fig6}.

\subsection{Low nematic order is porportional to the number of buds}

\begin{figure}[ht]

  \centering
  \includegraphics[width=0.7\textwidth]{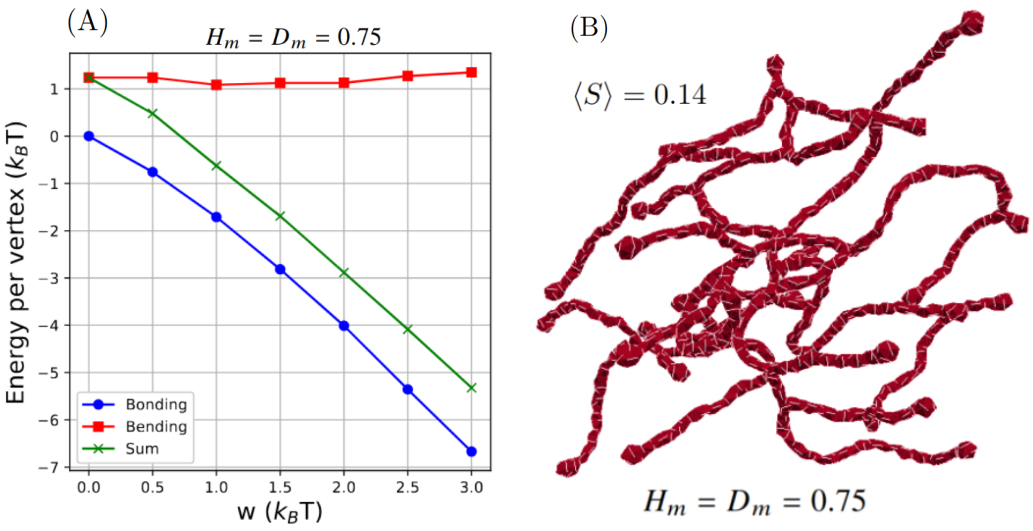}
  \caption{Total energy per vertex in dependence on $w$ for $H_m=D_m=0.75$ (A). Highly curved anisotropic inclusions result in very thin spaghetti-like shapes for normal interaction ($w=2$). The nematic order here is low because of the thin radius of the tubes (B).}
  \label{fig:fig12}
\end{figure}

Lastly we look at fully covered vesicles with nematic interaction with deviatoric curvature $H_m=D_m=0.75$ and all other parameters  same as before. The banana shape of the inclusion is preserved, but its curvature is increased. Figure \ref{fig:fig12}(A) shows the graphs of bending and bonding energy with respect to $w$. 

Bonding energy depends on two things; the strength of interaction $w$ and local nematic order. We find that high bonding energy (and hence worst nematic order)  goes to membranes with highest curved inclusions $H_m=0.75$.  This implies at least one of two things are taking place: either vesicles form many offshoot buds with their respective necks, or the vesicles resemble very elongated thin tubular structures. Figure \ref{fig:fig12}(B) shows such a final shape ($H_m=D_m=0.75$) and confirms our prediction of a low nematic order (and high bonding energy).

\section{Conclusion}
Membrane deformations in cells occur due to curved inclusions working in synergy with each other. Cell mechanics studied from a simple model with few assumtions might shed new light on formation of vesicle morphology. 

In this paper we demonstrated a new way of simulating anisotropic membrane inclusions by implementing the deviatoric elasticity model that was developed nearly twenty years ago \cite{iglivc2005role}. Although our investigation was limited only to banana membrane inclusions, we found that orientational order determines whether vesicles fully covered with inclusions result in pearls connected by necks or long tubes with few necks. Additionally, we observed that the average nematic order is inversely proportional to the number of thin necks formed by the vesicles.

The dynamics shown in Figure \ref{fig:fig4} may indicate a mechanism for primitive cell division or division of organelles inside the cell, using many curved banana-shaped membrane proteins. It is crucial that the membrane inclusions interact non-nematically. Nematic ordering of inclusions result in long narrow tubes and does not lead to fission-like geometries. This is a mechanism that will be explored in further papers.

An obvious advantage of this model is the absence of limitations of axial symmetry. On the other hand, the model has few parameters, and its phase space offers  further investigation. One could for example look at inclusion concentrations lower than $10\%$ or use a combination of curved inclusions. Some processes, like clathrin mediated endocytosis and formation of three way junctions in the endoplasmic reticulum \cite{chen2015lunapark,chen2012er}, are examples of membrane deformation and stabilisation of induced curvature due to a combination of curved inclusions. 

A further development would entail a constraint of constant volume. This could be achieved by introducing a new term to the Hamiltonian to include a volume potential to force the vesicles toward a wanted set volume. In this way, vesicle morphologies could be compared to the standard empty Helfrich diagram \cite{helfrich1973} and study the effects that inclusions have on the overall morphology. Furthermore, membranes with active inclusions could be modelled by adding a non-conservative force acting perpendicular away from the vesicle to simulate the polymerization of actin \cite{fosnaric2019theoretical,ravid2023theoretical}.

\section{Acknowledgements}
This work was supported by the Slovenian Research Agency (ARIS) through Grants  Nos.  J3-3066  and J2-4447  and Programme No.  P2-0232. N. G. is the incumbent of the Lee and William Abramowitz Professorial Chair of Biophysics, and acknowledges support by the Ben May Center for Theoryand Computation, and the Israel Science Foundation (Grant No. 207/22).

\section{Appendix - Methods}

\subsection{Monte-Carlo procedure}\label{app:procedure}

The membrane is represented by a set of $N$ vertices that are linked by tethers of variable length $l$ to form a closed, dynamically triangulated, self-avoiding two-dimensional network of approximately 2N triangles and with the topology of a sphere \cite{gompper1996,gompper2004}. The lengths of the tethers can vary between a minimal and a maximal value, $l_{min}$, and $l_{max}$, respectively. Self-avoidance of the network is ensured by choosing the appropriate values for $l_{max}$ and the maximal displacement of the vertex $s$ in a single updating step.

One Monte-Carlo sweep (MCs) consists of individual attempts to displace each of the $N$ vertices by a random increment in the sphere with radius $s$, centered at the vertex, followed by $R_B N$ attempts to flip a randomly chosen bond. We denote $R_B$ as the bond-flip ratio, which defines how many attempts to flip a bond are made per one attempt to move a vertex in one MCs. Note that the bond-flip ratio is connected to the lateral diffusion coefficient within the membrane, i.e. to the membrane viscosity. In this work we have chosen $R_B = 3$,  $s/l_{min}=0.15$ and $l_{max}/l_{min}=1.7$. The dynamically triangulated network acquires its lateral fluidity from a bond flip mechanism. A single bond-flip involves the four vertices of two neighboring triangles. The tether connecting the two vertices in diagonal direction is cut and reestablished between the other two, previously unconnected, vertices. The self-avoidance of the network is implemented by ensuring that no vertex can penetrate through the triangular network and that no bond can cut through another bond \cite{samo2015,fovsnarivc2019}.

\subsection{Anisotropic code details}
\subsubsection{Representing the membrane as a mesh}\label{app:code details}

Our solver is called \texttt{Trisurf}. It models the vesicle as a closed, triangulated surface: a graph with vertices $i\in V$ and edges $e_{ij}\in E$, and an auxiliary set of triangles $t_{ijk}\in T$. The triangles make the approximation of the surface, but it is the vertices which are the principle dynamical entities which holds the properties of the membrane (intrinsic curvature, membrane composition, nematic director, etc.) and move in space. 

From the position of the vertices, $\vec{x}_i $ we can compute the normal vector to each triangle $\vec{N}_{ijk}$ and circumcenter $\vec{O}_{ijk}$ (center of the circle containing the position of $\vec{x}_i,\vec{x}_j,\vec{x}_k$). This allows us to divide the triangle to six parts and assign two pieces to each vertex.

For vertex $i$, these are the sub-triangle between the vertex position $\vec{x}_i$, the triangle center $\vec{O}_{ijk}$, and the middle of one of the edges $\frac{\vec{x}_i+\vec{x}_j}{2}$, and a similar sub-triangle for the other edge $ik$. We can denote the vector for the side between the vertex and the edge middle as half the edge length $\vec{e}_{ij} =\vec{x}_j-\vec{x}_i$ and the side between center and the edge middle, which is half of the dual (voronoi) edge, as $\vec{\sigma}^{i}_{jk}$ (the other half of the voronoi edge is on the neighboring triangle $\sigma^{i}_{\ell j}$). We can compute this $\sigma$ from the circumcenter and the middle of $\vec{x}_i,\vec{x}_j$:
\begin{equation}\label{eq:VoronoiEdge}
    \vec{\sigma}^{i}_{jk} = \frac{\vec{x}_i+\vec{x}_j}{2} - \vec{O}_{ijk} 
\end{equation}

With this, we assign an area $A\left(i\right)$ and a normal $N\left(i\right)$ for each vertex $i$, by running over the neighbors $j-1,j,j+1\ldots$ and the adjacent triangles $\left(i,j-1,j\right),\left(i,j,j+1\right)\ldots$
\begin{equation} \label{eq:VertexArea}
    A\left(i\right) = \sum_{\left\langle i,j\right\rangle} \frac{1}{2} \frac{\left|e_{ij}\right|}{2}\left|\sigma^i_{j,j+1}\right| + \frac{1}{2} \frac{\left|e_{ij}\right|}{2}\left|\sigma^i_{j,j-1}\right|
\end{equation}
\begin{equation} \label{eq:VertexNormal}
    \vec{N}\left(i\right) = \frac{\sum_{\left\langle i,j\right\rangle}  \vec{N}_{ij,j+1} \left|e_{ij}\right|\left|\sigma^i_{j,j+1}\right| +  \vec{N}_{i,j-1,j} \left|e_{ij}\right|\left|\sigma^i_{j,j-1}\right|}{\left|\ldots\right|}
\end{equation}
Where the $j$ vertices are the counterclockwise ordered neighbors of $i$.

The fluidity of the surface is achieved by bond-flips, where a bond $ij$ and the two triangles that share it is $ijk,ji\ell$ are replaced by a cross bond $k\ell$ and two triangles $i\ell k,j k \ell$, which allows vertices to change neighbors.

\subsubsection{Anisotropic curvature on the vesicle}

To get the anisotropic bending energy of the surface, we use the method by \cite{ramakrishnan2010monte} to estimate the shape operator matrix $S$ on each vertex, which represents the shape of the surface at the point. We then calculate the mismatch tensor $M=S-C_m$, where $C_m$ is the intrinsic curvature tensor whose direction is determined by the director and the other tangent vector to the local normal ($\hat{t}=\hat{N}\times\hat{d}$)
\begin{equation}
    C_m = \frac{C_0+D_0}{2}\hat{d}\otimes\hat{d} + \frac{C_0-D_0}{2} \hat{t}\otimes\hat{t}
\end{equation}
where $C_0,D_0$ are the spontaneous curvature and spontaneous deviator at the vertex, respectively, which reflects the physical characteristics of local membrane composition.

The bending energy is calculated by inserting the mismatch tensor in the Hamiltonian
\begin{equation}\label{eq:eq13}
    E_1=\frac{K_1}{2}\left(\mathrm{Tr} M\right)^2 + K_2 \mathrm{Det}M
\end{equation}
Where $K_1$ and $K_2$ are the bending moduli of the vertex, again reflecting physical parameters due to local composition.

To calculate the shape curvature of a vertex $i$ based on \cite{ramakrishnan2010monte}, each edge $ij$ is assigned a shape tensor estimation
\begin{equation}
    S_{ij}=h_{ij} \; \vec{b}\otimes \vec{b}
\end{equation}
$\vec{b}$ is the binormal at the edge $\hat{N}_{ij}\times \vec{e}_{ij}$, where $\vec{e}_{ij} = \vec{x}_j-\vec{x}_i$ is the edge vector and $\hat{N}_{ij}$ is the normal of the edge $\hat{N}_{ij} = \left(\hat{N}_{i,j-1,j} + \hat{N}_{i,j,j+1} \right)/ \left|\ldots\right| $ which is the sum of the normal of the two triangles sharing the edge, normalized.
$h_{ij}$ is a factor representing the directional derivative of the area $\approx \nabla_p A$
\begin{equation}
    h_{ij} = 2 \left|e_{ij}\right|\cos\left(\frac{\Phi}{2}\right)
\end{equation}
Where $\Phi$ is the dihedral angle (angle between the two triangle sharing the edge).
Luckily there is a simple triple product formula for this factor
\begin{equation}
    h_{ij} = 2\left|\vec{e}_{ij}\right| \vec{N}_{ij} \cdot \left(\vec{N}_{i,j-1,j} \times \hat{e}_{ij} \right)
\end{equation}
The cross product of the edge direction and a triangle normal gives a vector on the triangle which is perpendicular to the edge, which is at an angle $\frac{\Phi}{2}$ from the edge normal. 

The full vertex shape tensor is a sum of the edge tensor, weighted by the match of the normal of the edge to the normal of the vertex.
\begin{equation}
    S\left(i\right) = \frac{1}{A\left(i\right)}\sum_j \hat{N}\left(i\right)\cdot\hat{N}_{ij} \; S_{ij}
\end{equation}

We then project this 3x3 matrix in real space $\hat{x},\hat{y},\hat{z}$ to the tangent plane of the vertex $\hat{d},\hat{t}$
\begin{equation}
    S_{2\times2} =\begin{pmatrix}
        \hat{d}\cdot S \cdot \hat{d} & \hat{d} \cdot S \cdot \hat{t} \\
        \hat{t}\cdot S \cdot \hat{d} & \hat{t} \cdot S \cdot \hat{t}
    \end{pmatrix}
\end{equation}

The mean curvature $H$ at the vertex is half the trace, while the Gaussian curvature $K$ is the determinant, which are two of the degrees of freedom in the Hamiltonian.

The mismatch matrix can be calculated
\begin{equation}
    M = \begin{pmatrix}
        S_{dd} - \frac{C_0 + D_0}{2}& S_{dt} \\
        S_{td} & S_{tt} -  \frac{C_0 - D_0}{2}
    \end{pmatrix}
\end{equation}
The angle between the director and the eigenvectors of the shape matrix is what results in the $\omega$ angle dependence, which is the final degree of freedom.
The mismatch tensor is simply plugged to the equation \ref{eq:eq13} to compute the bending energy of the vertex. If we integrate it across the whole membrane, we get equation \ref{eq:eq1}.

\subsubsection{Dicretization of the order parameter}\label{app:order}

The order parameter is calculated by a python script shown below. The director gives the direction of the inclusion $\vec{n}$.

\begin{lstlisting}
import numpy as np
import sys
sys.path.append('path_to_vtu_file')
import vtu

v=vtu.PyVtu('timestep_x.vtu')

nvtx=len(v.type)
nematic_order=np.zeros(nvtx)

for i in np.arange(nvtx)[v.type!=4]:
    js = v.get_neighbors(i)
    js = js[v.c0[js]>0]
    di = v.director[i]
    nematic_order[i] = np.nanmean(3*(v.director[js]@di)**2-1)/2

v.update_array('nematic_order',nematic_order)
v.update_all()
v.write_vtu('nematic_order.vtu')
\end{lstlisting}

\bibliographystyle{unsrt}  
\bibliography{references}  

\end{document}